\documentstyle[12pt,aaspp4,flushrt]{article}

\newcommand\psra{B1046$-$58}
\newcommand\psrb{J1105$-$6107}
\newcommand\groa{2EG J1049$-$5847}
\newcommand\grob{2EG J1103$-$6106}

\newcommand\egret{EGRET}
\newcommand\cgro{{\it Compton Gamma-Ray Observatory}}
\newcommand\gray{$\gamma$-ray}
\newcommand\grays{$\gamma$-rays}
\newcommand\lapp{\ifmmode\stackrel{<}{_{\sim}}\else$\stackrel{<}{_{\sim}}$\fi}
\newcommand\gapp{\ifmmode\stackrel{>}{_{\sim}}\else$\stackrel{>}{_{\sim}}$\fi}

\begin{document}

\title{High-Energy Gamma-Ray Observations of Two Young, Energetic Radio Pulsars}

\author{V. M. Kaspi\altaffilmark{1}  and J. R. Lackey\altaffilmark{2}}
\affil{Department of Physics and Center for Space Research, 
Massachusetts Institute of Technology, 70 Vassar St.,
Cambridge, MA 02139}

\author{J. Mattox\altaffilmark{3}}
\affil{Astronomy Department, Boston University, 725 Commonwealth
Ave., Boston, MA 02215}

\author{R. N. Manchester\altaffilmark{4}}
\affil{Australia Telescope National Facility, Epping, NSW, 2121 Australia}

\author{M. Bailes\altaffilmark{5}}
\affil{Astrophysics and Supercomputing, Swinburne University of Technology, Mail 31, PO Box 218, Hawthorn, Victoria, 3122 Australia}

\author{R. Pace\altaffilmark{6}}
\affil{Physics Department, University of Adelaide, Adelaide, SA, Australia}

\altaffiltext{1}{Alfred P. Sloan Research Fellow; vicky@space.mit.edu}
\altaffiltext{2}{jes@space.mit.edu}
\altaffiltext{3}{mattox@bu.edu}
\altaffiltext{4}{rmanches@atnf.csiro.au}
\altaffiltext{5}{mbailes@swin.edu.au}


\begin{abstract}

We present results of \cgro/\egret\ \linebreak observations of the
unidentified high-energy $\gamma$-ray sources \groa\ (GEV
J1047$-$5840, 3EG J1048$-$5840) and \grob\ (3EG J1102$-$6103).  These
sources are spatially coincident with the young, energetic radio
pulsars PSRs \psra\ and \psrb, respectively.  We find evidence for an
association between PSR~\psra\ and \groa.  The $\gamma$-ray pulse
profile, obtained by folding time-tagged photons having energies above
400~MeV using contemporaneous radio ephemerides, has probability of
arising by chance of $1.2 \times 10^{-4}$ according to the
binning-independent H-test.  A spatial analysis of the on-pulse
photons reveals a point source of equivalent significance
$10.2\sigma$.  Off-pulse, the significance drops to $5.8\sigma$.
Archival {\it ASCA} data show that the only hard X-ray
point source in the 95\% confidence error box of the \gray\ source is
spatially coincident with the pulsar within the 1$'$ uncertainty
(\cite{pkg99}).  The double peaked \gray\ pulse morphology and leading
radio pulse are similar to those seen for other \gray\ pulsars
and are well-explained in models in which the \gray\ emission is
produced in charge-depleted gaps in the outer magnetosphere.  The
inferred pulsed \gray\ flux above 400~MeV, $(2.5 \pm 0.6) \times
10^{-10}$~erg~cm$^{-2}$~s$^{-1}$, represents 0.011$\pm$0.003 of the
pulsar's spin-down luminosity, for a distance of 3~kpc and 1~sr
beaming.  For PSR~\psrb, light curves obtained by folding \egret\
photons using contemporaneous radio ephemerides show no significant
features.  We conclude that this pulsar converts less than 0.014 of
its spin-down luminosity into $E>100$~MeV \grays\ beaming in our
direction (99\% confidence), assuming a distance of 7~kpc, 1~sr
beaming and a duty cycle of 0.5.

\end{abstract}


\section{Introduction} \label{sec:intro}

An outstanding unknown in \gray\ astronomy is the nature of a
population of high-energy \gray\ sources at low Galactic latitude
which were first identified using the COS-B satellite (\cite{sbb+81}).
More recently, these sources were studied in greater detail by the Energetic
Gamma Ray Experiment Telescope (\egret) aboard the {\it Compton
Gamma-Ray Observatory} (\cite{kbd+96}).  In contrast to the
high-energy \gray\ sources at high latitudes, which are typically
associated with active galaxies, several dozen apparent point sources
of high-energy $\gamma$-rays at latitudes $|b| \lapp 10^{\circ}$ have as
yet not been identified with astronomical objects.  However, a handful
of low-latitude sources has been identified with young, energetic
radio pulsars (see \cite{tho96b} for a review).  This suggests that
many of the unidentified sources are also young pulsars.  This
hypothesis has been considered in various studies, many of which
suggest that the majority, if not all, of the unidentified sources are
pulsars (\cite{yr95,kc96,yr97,cz98}).  Sturner \& Dermer (1995)
\nocite{sd95,ehks96} suggested that many of the sources are not
pulsars but rather are a result of cosmic ray interactions with gas in
supernova remnants.  With the \egret\ spatial resolution generally limited to
tens of arcminutes, identification of discrete
high-energy $\gamma$-ray sources is difficult using spatial coincidence alone,
explaining why this problem has remained open so long.

Rotation-powered pulsars permit a straightforward determination of the
nature of a \gray\ source using the signature pulsed emission.  Indeed
the Crab and Vela pulsars are long-known to be strong sources of
pulsed $\gamma$-rays.  \egret\ has provided unambiguous and/or strong
evidence for pulsations from five other rotation-powered pulsars (see
Table~\ref{ta:psrs}).  Five of the six pulsars having
the highest values of the parameter $\dot{E}/d^2$, where the spin-down
luminosity $\dot{E} \equiv 4\pi^2 I \dot{P}/P^3$ ($P$ is the
pulse period and $\dot{P}$ its rate of increase) and $d$ is the
distance to the source, are established sources of high-energy $\gamma$-ray
pulsations.  Finding rotation-powered pulsars at
\gray\ energies is important because those detected thus far have their
maximum luminosities in this energy range.  Thus \grays\ may hold the
key to the poorly understood pulsar energy budget and emission
mechanism, in which mechanical energy of rotation is efficiently
converted into non-thermal high-energy radiation.  The origin of the
\grays\ -- whether at the neutron star polar cap (e.g. \cite{dh96}) or
in the outer magnetosphere (e.g. \cite{ry95}) -- is still debated.  In
this paper we examine high-energy \gray\ data from the direction of
two young, energetic pulsars which are good candidates for observable
\gray\ pulsations.

PSR~\psra\ is a 124~ms radio pulsar having characteristic age 20~kyr
and spin-down luminosity $\dot{E}=2.0 \times 10^{36}$~erg~s$^{-1}$.  It
was discovered in a survey of the Galactic plane for radio pulsars
(\cite{jlm+92}), and has an unremarkable radio average pulse profile,
characterized by a single narrow peak having width $\sim$8~ms at
1.4~GHz (Figure~\ref{fig:1046prof}).  Its pulsed radio emission is
highly linearly polarized and shows a slow position angle variation
across the pulse, suggestive of conal emission (\cite{qmlg95}).  The
dispersion measure of 129.01$\pm$0.01~pc~cm$^{-3}$ toward this pulsar
implies a distance of 2.98$\pm$0.35~kpc (\cite{tc93}).  In the
rank-ordered $\dot{E}/d^2$ list, PSR~\psra\ falls ninth, with six of
the eight above it known \gray\ pulsars, five of those six being
high-energy $\gamma$-ray sources  (see Table~\ref{ta:psrs}).
The pulsar's position lies near the 95\% confidence contour for the
Second EGRET catalog source \groa\ (\cite{tbd+95}).  Thompson et
al. (1994) \nocite{tab+94} reported a pulsed flux 99.9\% confidence
upper limit for energies above 100~MeV of $4.5 \times
10^{-7}$~photons~cm$^{-2}$~s$^{-1}$ for a 50\% duty cycle pulse.  They
noted, however, a ``hint'' of a pulsation for PSR~\psra\ above 1~GeV,
having probability of chance occurrence of 14\%.  As reported in a companion
paper, Pivovaroff, Kaspi \&
Gotthelf (1999) have detected X-rays from the direction of the pulsar;
that emission is not pulsed and is probably due to a synchrotron
nebula.

PSR~\psrb\ is a 63~ms pulsar having characteristic age 63~kyr and
$\dot{E}=2.5 \times 10^{36}$~erg~s$^{-1}$ (\cite{kbm+97}).  It
exhibits a narrow double peaked radio pulse profile having width
3.4~ms at 1.6~GHz.  Its radio emission is highly linearly polarized,
showing a slow positional angle variation suggestive of
conal emission (\cite{ckml99}).
The pulsar's dispersion measure of 270.55$\pm$0.14~pc~cm$^{-3}$ implies
a distance of 7.1$\pm$0.9~kpc (\cite{tc93}).  Ranking by $\dot{E}/d^2$, it is
23rd, well above the known \gray\ pulsar PSR~B1055$-$52 (whose
distance may well be overestimated; see \cite{tbb+99} and references
therein) though well below several sources which have not been
detected by EGRET (\cite{tab+94}).  PSR~\psrb\ is spatially coincident
with the unidentified EGRET source \grob\ (\cite{tbd+95,kbm+97}).  This \gray\
source has pulsar-like properties: it has a hard spectrum and is not
time variable.  However the Second EGRET catalog noted marginal
evidence for the source being extended.  PSR~\psrb\ has been detected as
an unpulsed X-ray source (\cite{gk98,skg99}); this emission is best
explained as being due to nebular synchrotron emission powered by the
pulsar wind.

Given the importance of identifying \gray\ pulsations in radio pulsars
as well as establishing the nature of the unidentified EGRET sources,
additional EGRET exposure of the Eta Carina region was obtained.  We
report here on an analysis of EGRET data for PSR~\psra\ obtained
between 1991 May and 1997 October, and for PSR~\psrb\ between 1993
July and 1997 October.  In \S\ref{sec:obs} we summarize the EGRET
observations used in our analysis.  In \S\ref{sec:results} we report
our findings, showing evidence for \gray\ pulsations from PSR~\psra,
but none from PSR~\psrb.  We discuss our results and their
implications in \S\ref{sec:discuss}.

\section{Observations and Analysis}  \label{sec:obs}

The Energetic Gamma Ray Experiment Telescope (EGRET) aboard the {\it
Compton Gamma Ray Observatory} is sensitive to photons in the energy
range from approximately 30~MeV to  30~GeV, using a multi-level spark
chamber system to detect electron-positron pairs resulting from \grays.  
A NaI calorimeter provides energy resolution of 20-25\%, and
a plastic scintillator anti-coincidence dome prevents triggering on
events not associated with high-energy photons.  The detector thus has
negligible background.  The detector is described in more detail in
Thompson et al. (1993) \nocite{tbf+93a} and references therein.  The
data sets used in this analysis consist of archival data, as well as PI
data for Cycles 5 and 6.  Data sets included event UTC arrival times
accurate to 100~$\mu$s, energy, measured direction of origin, satellite
location, and other information.  A summary of the observations of
PSRs~\psra\ and \psrb\ is presented in Tables~\ref{ta:1046obs} and
\ref{ta:1105obs}.

\subsection{Timing Analysis}

The timing analysis was done using the interactive software {\tt
PULSAR} which has been described elsewhere (e.g. \cite{fie95}) and is
based on the {\tt TEMPO} software package (e.g. \cite{tw89}).  Event
arrival times are first reduced to the solar system barycenter and
then folded using a contemporaneous ephemeris derived from radio
monitoring.

Young pulsars are notorious for displaying timing irregularities,
namely glitches and long-term timing ``noise'' (see \cite{lyn96} for
a review).  The pulsars in question are no exception
(\cite{jml+95,kbm+97}).  Timing noise alone can result in phase
deviations as large as a full cycle over a year or two, while glitches
can result in so large a phase deviation that only unusually dense
observations can permit unambiguous phase determination.  Such
deviations cannot be tolerated when searching for \gray\ pulsations
over many years, particularly when expecting weak pulsations, as even
small deviations could broaden the
signal beyond detectability.  

For this reason, we have used contemporaneous radio ephemerides
determined by timing observations made at the 64-m radio telescope at
Parkes, Australia.  Most radio observations were done at a central
frequency near 1500~MHz, with some at 430 and 660 MHz.  Prior to 1995,
all data were obtained using filter-bank timing systems (2 $\times$ 64
$\times$ 5~MHz at 1520~MHz and 2 $\times$ 256 $\times$ 0.125~MHz at
430 and 660~MHz) that have been described elsewhere
(e.g. \cite{bhl+94}).  Most data from after 1995 were obtained using
the Caltech correlator-based pulsar timing machine (\cite{nav94}),
which has $2 \times 128$ lags across 128~MHz in each of two separate
frequency bands.  Typically, correlator observations were made at
central frequencies of 1420 and 1650~MHz simultaneously.  Filter-bank
data were recorded on tape and folded off-line; correlator data were
folded on-line.  Resulting folded profiles were convolved with high
signal-to-noise templates to yield topocentric pulse arrival times.
Resulting arrival times were analyzed using the standard {\tt TEMPO}
pulsar timing software package\footnote{{\tt
http://pulsar.princeton.edu/tempo}} together with the JPL~DE200
ephemeris (\cite{sta82}).  Ephemerides for folding the \gray\ data
were determined piecewise using a minimum of some two dozen arrival
times per valid interval, with RMS residuals well under the \gray\
binning resolution in all cases.  The ephemerides used are provided
for each EGRET viewing period in publicly accessible machine-readable format at
{\tt http://www.atnf.csiro.au/research/pulsar/psr/archive/}.

Since the range of radio frequencies observed at each epoch was not
generally sufficiently large to determine dispersion measures with
high precision, we assume that the dispersion measures do not vary
significantly.  Substantial variations would result in changes in the
pulse phase extrapolated from the radio ephemeris to the effectively
infinite \gray\ frequency according to the cold plasma dispersion law.
This could cause smearing of the pulse profile.  We can test the
plausibility of our assumption of a constant dispersion measure (DM)
using the results of Backer et al. (1993) \nocite{bhvf93} who
empirically characterized variations in DM for several pulsars.  As
neither pulsar considered here lies within a detectable supernova
remnant (\cite{jml+95,kbm+96,gk98,pkg99}), the very large DM
variations observed for the Crab and Vela pulsars, usually attributed
to their unusual environments, are unlikely to be present.  From the
empirical findings of Backer et al. (1993), for PSR~\psra, the maximum
DM variation per year expected is $\sim$0.01~pc~cm$^{-3}$.  Over the
$\sim$8~yr duration of the experiment described here, this amounts to
a variation in pulse phase of $\sim$1.2$\times 10^{-3}$, a factor of
$\sim$40 smaller than our binning resolution.  For PSR~\psrb, the DM
variation could be an order of magnitude larger, however the
experiment in this case spans only $\sim$4~yr.  The phase variation
then could be as large as $\sim$1.1$\times 10^{-2}$, smaller, but only
by a factor of $\sim$4, than our binning resolution.  We conclude that
DM variations are certainly negligible for PSR~\psra, and probably
negligible for PSR~\psrb.

\subsection{Spatial Analysis}

\label{sec:spatial}

The energy dependence of the  point-spread function of the EGRET
detector can be characterized by the function
\begin{equation}
\theta_{67}(E) = 5^{\circ}.85 \times (E/100)^{-0.534}, 
\label{eq:theta} 
\end{equation}
where $\theta_{67}$ is the half-angle of the cone encircling the
actual source direction and which contains 67\% of the events
having energy $E$ in MeV (\cite{tbf+93a}).  
In our timing analysis, we used all events
within $\theta_{67}$.  The source directions are
(J2000) RA 10$^{\rm h}$~48$^{\rm m}$~12$^{\rm s}$.6(3),
DEC $-$58$^{\circ}$~32$^{\prime}$~03$^{\prime\prime}$.75(1)
  for PSR~\psra\ (\cite{sgjf99}), and
(J2000) RA 11$^{\rm h}$~05$^{\rm m}$~26$^{\rm s}$.07(7),
DEC $-$61$^{\circ}$~07$^{\prime}$~52$^{\prime\prime}.1(4)$
for PSR~\psrb\ (\cite{kbm+97}), where the numbers in brackets
are $1\sigma$ 
uncertainties.
The total number of counts folded is given
for each viewing period in Tables~\ref{ta:1046obs} and
\ref{ta:1105obs} for PSRs~\psra\ and \psrb, respectively.

Spatial analysis of \gray\ data is done using the likelihood ratio
test
(\cite{mbc+96}).  The significance of a point source is found using
a test statistic $T_s$ that compares the likelihood  of the distribution
of  counts (which is governed by Poisson statistics) having occurred
under the null hypothesis of no point source, with that with a point
source present, given the instrumental point-spread function (PSF).
Mattox et al. have shown that a point-source significance in units
of the standard deviation is given by $\sqrt{T_s}$ where
\begin{equation}
T_s \equiv 2(\ln L_1 - \ln L_0),
\end{equation}
where $L_1$ is the likelihood under the assumption of a point
source and $L_0$ is that under the null hypothesis.  The
likelihood itself is given by
\begin{equation}
L = \prod_{ij} \frac{\theta_{ij}^{n_ij} e^{-\theta_{ij}}}{n_{ij}!},
\end{equation}
where the product is over pixel $ij$ that contains $n_{ij}$ counts,
and the model prediction is $\theta_{ij}$.

\section{Results} \label{sec:results}

\subsection{PSR~\psra} \label{sec:1046}

\subsubsection{Timing Analysis} \label{sec:1046time}

\begin{figure}
\plotone{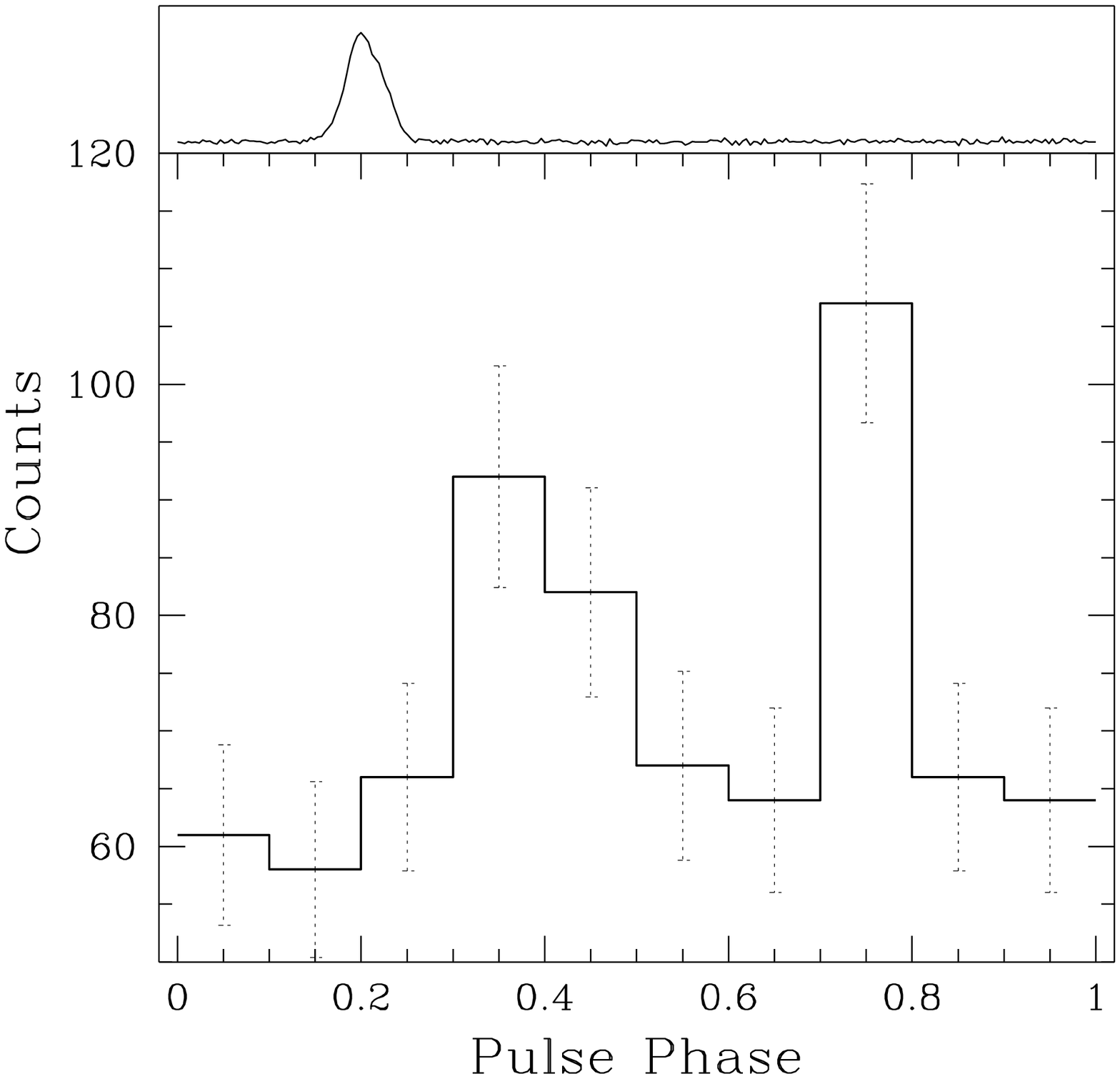}
\caption{Bottom panel: folded EGRET light curve for PSR~B1046$-$58
for $E>400$~MeV.  Top panel: radio profile at 20~cm from the Parkes
observatory.}
\label{fig:1046prof}
\end{figure}

In Figure~\ref{fig:1046prof}, we present the folded light curve for
PSR~\psra\ for energies above 400~MeV.  The significance of the
apparent modulation can only be established using appropriate
statistical tests.  The optimal test is the H-test (\cite{dej94}),
which provides a figure of merit for a folded light curve that is
independent of the number of bins or of the phase of the putative
signal.  For this reason, the H-test is superior to the standard
$\chi^2$ test (e.g. \cite{lde+83}).  Further, the H statistic has been
shown to be optimal when considering a light curve having unknown
pulse morphology.  In this sense, it is superior to the $Z_N^2$ test
(\cite{bbb+83}) which can be done for specified numbers of harmonics
only.  For the light curve in Figure~\ref{fig:1046prof}, the H
statistic is 22.5, which has a probability of $1.2 \times 10^{-4}$ of
having occurred by random chance, with 5 harmonics indicated.  This
corresponds to an equivalent normal distribution deviation of
$3.9\sigma$ from the null hypothesis of no pulsations.  For reference,
$\chi^2 = 26.3$ for 10 bins (9 degrees of freedom); the probability of
this value or higher having occurred by chance is $1.8 \times 10^{-3}$
(3.1$\sigma$).  The $Z_4^2$ statistic has value 33.8; the probability
of this value or higher having occurred by chance is $4.3 \times
10^{-5}$ (4.1$\sigma$).  Note that there is effectively only a single
trial folding period.  Since we considered two pulsars, these
probabilities should be multiplied by a factor of two.  Although we
did not search independent energy bands, the choice of 400~MeV lower
energy cutoff was made to optimize the significance; we discuss this
further below.

We have checked the robustness of our result in several ways.  First,
we divided the total data set into two halves of approximately equal
size and verified that the signal is seen in both, albeit at reduced
significance, as expected.  For the first half data set (MJDs
48386--49361), we find H=16.7, which has probability
$1.3\times10^{-3}$, while for the second half data set (MJDs
49361--50728) we found H=15.0, having probability $2.4\times 10^{-3}$.
These statistics are for $E>400$~MeV.  Second, we verified that the
signal persisted when the energy-dependent cone opening angle
(Eq.~\ref{eq:theta}) was increased to include 99\% of
the counts from the known source direction; in this case
we found H=12.5, which has probability $6.8\times 10^{-3}$.

Note that zero is suppressed in the light curve shown in 
Figure~\ref{fig:1046prof}.
The observed pulsed
fraction for $E>400$~MeV is $0.12 \pm 0.02$.  The off-pulse
\grays\ may originate as diffuse Galactic emission --- the pulsar
is located in the complicated Eta Carina region, tangent to the Carina
spiral arm, which is known to contain significant diffuse emission and
many discrete sources.
(\cite{bdf+93,hbc+97}).  In \S\ref{sec:1046spatialres} we describe a
spatial analysis that indicates that a small fraction of the off-pulse
could be emission from PSR~B1046$-$58 itself.

\begin{figure}
\plotone{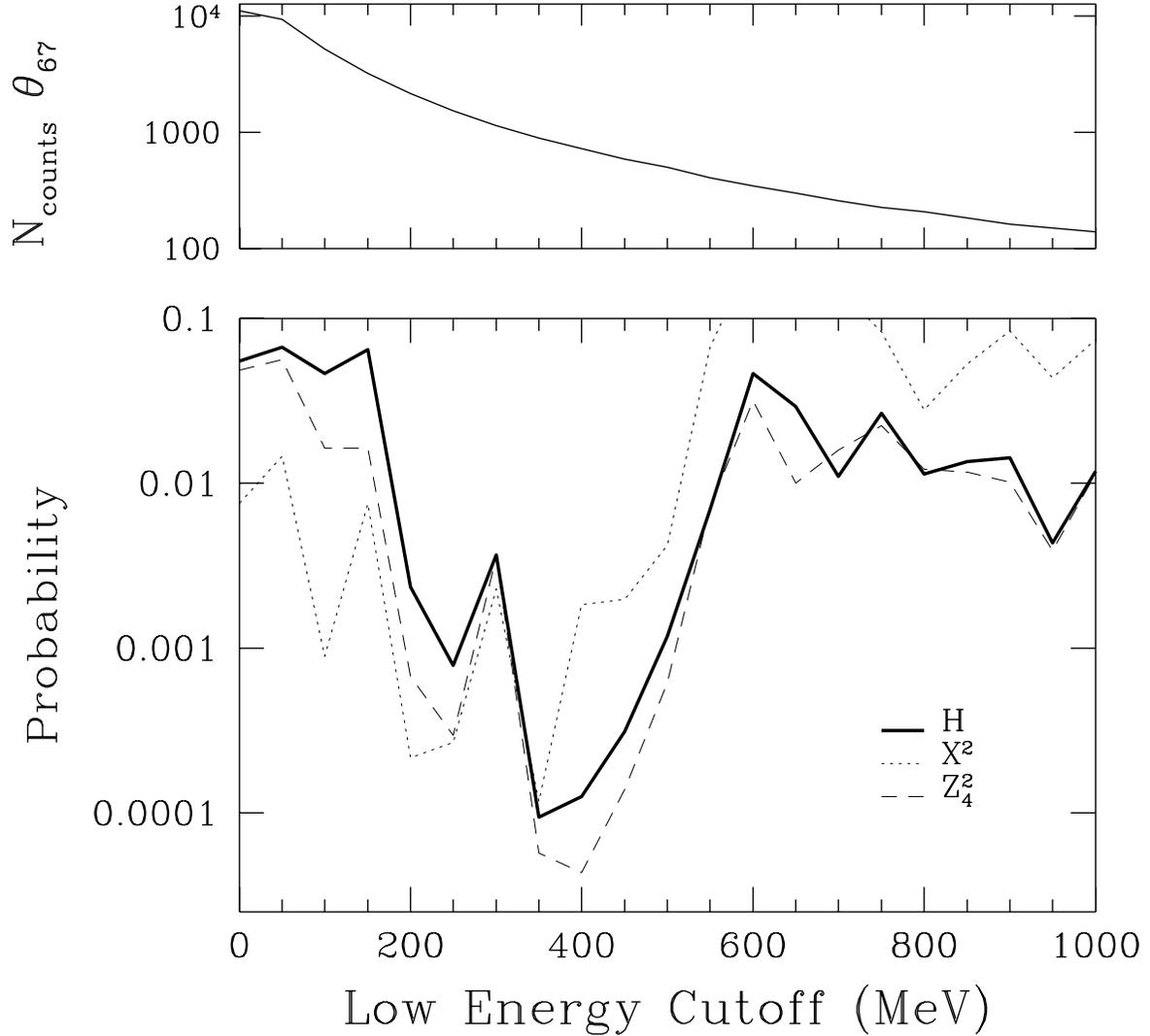}
\caption{Upper panel: number of events within $\theta_{67}$ for
PSR~B1046$-$58 as a function of lower energy cutoff in MeV.  Lower
panel: results of three statistical tests applied to folded profiles
for PSR~B1046$-$58 as a function of lower energy cutoff in MeV.  The
probabilities represent the chance of the measured value of the
statistical test or higher having been obtained by chance.  The three
tests are the H test, the $\chi^2$ test for 10 bins (9 degrees of
freedom), and the $Z_4^2$ test.}
\label{fig:1046stats}
\end{figure}

One way to check for the presence of contaminating diffuse emission is
to utilize the different expected spectra of diffuse and pulsed
signals.  In Figure~\ref{fig:1046stats} we show, in the lower panel,
results of three statistical tests
as a function of lower cutoff energy.  The probability shown is that of
finding the value of the statistical test, or higher, for a profile 
consisting of folded photons having energies above the cutoff.  The upper
panel shows the number of photons included in the analysis, here done
for counts within $\theta_{67}$.  Qualitatively, the behavior is as
expected: the diffuse emission has a softer spectrum than those of the
known \gray\ pulsars, so as the lower energy bound is increased,
the significance of the pulse should increase until the pulsar signal is
too photon-starved to be detected.
Furthermore, the EGRET PSF narrows with
increasing energy (Eq.~\ref{eq:theta}), so diffuse emission and nearby
confusing point sources are preferentially removed as the lower energy
bound is increased given advance knowledge of the pulsar position
(\cite{lm97}).  However, the sudden decrease in significance when the lower
energy cut changes from 400 to 600~MeV is unexpected, given the
smooth decrease in the number of counts.  This could indicate
a spectral cutoff in the pulsed emission, or could simply be a
result of the poor statistics for the pulse profile; it could also
suggest the apparently significant modulation is a statistical
fluctuation.

The detection of pulsed high-energy \gray\ emission from PSR~B1951+32
by Ramanamurthy et al. (1995) \nocite{rbd+95} was done by considering
events selected from a cone of fixed angle with respect to the pulsar
position, as a compromise between maximizing signal from the source
and minimizing background contamination.  We have carried out the same
analysis for PSR~B1046$-$58, using a fixed cone of $2^{\circ}$
half-angle.  We chose this size cone as it resulted in a comparable
total number of events as for the energy-dependent cone.  For the
fixed cone, we find H$=23.9$, corresponding to a probability of $6.9
\times 10^{-5}$ (4.0$\sigma$), an improvement over the
energy-dependent-cone results.  This supports the reality of the
pulsations.  However, not all the profile statistics improve for the
fixed cone; for example, the $\chi^2$ statistic drops to 25.3,
corresponding to a probability of $2.7 \times 10^{-3}$ (3.0$\sigma$),
still statistically significant, but less so than with the
energy-dependent cone.

Ramanamurthy et al. (1996) \nocite{rfk+96} reported high-energy \gray\
pulsations from the radio pulsar PSR~B0656+14 using photons
that had been ``weighted'' according to their energy and direction,
and the known telescope response.  We have employed a similar algorithm to
theirs in considering the apparent modulation seen in the unweighted
data.  Each event is assigned a weight which is equal to the
probability that a specific \gray\ originated from the point source
(rather than as diffuse Galactic emission),
\begin{equation}
w= {{\rm PSF}(r,E)\over {\rm PSF}(r,E) + {B\over S}(E)},  
\end{equation}
where PSF$(r,E)$ is
the energy-dependent EGRET point spread function, the
probability per steradian for a photon of energy $E$ to be detected at a
measured angle $r$ from the source.  ${B/S}$ is the ratio of diffuse
Galactic counts per steradian to source counts at energy $E$.  As the
combinations of weights within phase bins are, unlike the unweighted
data, not described by Poisson statistics, the same statistical tests
do not apply.  Instead, we considered the simple statistical standard
deviation of the binned values, $x_i$, $\sum_{i=1}^{N} (x_i -
\overline{x})^2/(N-1)$.  We used Monte Carlo simulations of profiles
to determine the probability of chance occurrence of a profile having
this statistic.  Simulated profiles of weighted data were made by
randomly varying the phases.  In this way, for $E > 400$~MeV, we found
that the probability of obtaining a profile having the same modified
$\chi^2$ statistic or higher for the folded, weighted data was 0.026.
Thus the significance of the detection decreases using the
weights. This was not expected.  However, given the evidence below in
favor of the association, it does not eliminate the
possibility that pulsed flux is being detected from PSR~B1046$-$58;
the decrease in significance could be due to the poorly known
pulsed \gray\ spectrum (which was assumed to be a power law
of index 1.97$\pm$0.07 -- see \S\ref{sec:spectrum}) or to a statistical
fluctuation.

\subsubsection{Spatial Analysis}
\label{sec:1046spatialres}

We have conducted a likelihood analysis (\cite{mbc+96}) of the EGRET
data for PSR~\psra\ (see \S\ref{sec:spatial}) for $E>400$~MeV events.
In the analysis, the model for the diffuse Galactic flux of 
Hunter et al. (1997)
\nocite{hbc+97} was used, and the flux of nearby sources in the Second
EGRET catalog (\cite{tbd+95}) was estimated simultaneously with that
from the position of PSR~\psra.  A point source at the pulsar position
was detected with a significance of $\sqrt{T_s}=10.8\sigma$.  We
estimate 347$\pm$40 $\gamma$-ray events from this point source,
corresponding to a flux of (1.6 $\pm$
0.2)$\times10^{-7}$~photons~cm$^{-2}$s$^{-1}$ ($E> 400$~MeV).  The
distribution of EGRET events is consistent with the EGRET PSF; we find
a reduced $\chi^2$ of 1.3 for the sum of all events for Cycles 1--6 of
the EGRET mission.

A similar analysis was done for different rotational phases.
Phase-resolved event maps for $E>400$~MeV were made for peak 1 (phases
0.300--0.475), interpeak 1 (0.475--0.675), peak 2 (0.675--0.825), and
interpeak 2 (0.825--0.300).  The peak 1 and peak 2 maps were combined
to form an ``on-pulse'' map.  The interpeak 1 and interpeak 2 maps
were combined to form an ``off-pulse'' map.

For the on-pulse map, a point source at the position of PSR~\psra\ was
detected with a significance of $\sqrt{T_s}=10.2\sigma$, corresponding
to an estimated 203$\pm$25 $\gamma$-ray events.  For the off-pulse map, a
point source at the position of PSR~\psra\ was detected with a
significance of $\sqrt{T_s}=5.8\sigma$ corresponding to an estimated
144$\pm$29 $\gamma$-ray events.  The distribution of EGRET events was
consistent with the PSF for both the on- and off-pulse phase
selections. The EGRET position estimate was also consistent with that
of PSR~\psra\ for both the on- and off-pulse phase selections.  Thus,
significant flux is detected during both the on- and off-pulse phases,
and it appears that both the pulsar and diffuse Galactic $\gamma$-ray
emission contribute to the emission represented by the suppressed zero in
Figure~\ref{fig:1046prof}.

We also subdivided the on-pulse map into two, by making separate maps
for peak 1 and peak 2 data.  Both peak 1 and peak 2 show a point
source at the position of the pulsar, with significances of 6.0$\sigma$ and
8.4$\sigma$, respectively.

Maps of ``residual'' flux were also made for both the on- and off-pulse
phase selections by subtracting the counts distribution for nearby
sources, counts corresponding to an isotropic flux of
$0.295\times10^{-5}$~photons~cm$^{-2}$~s$^{-1}$~sr$^{-1}$, and counts
corresponding to diffuse Galactic flux (with an empirically determined
scaling factor of 0.242 for the model of Hunter et al. 1997
\nocite{hbc+97} for $E> 100$~MeV flux).  The residual counts map was
then converted to a residual flux map by dividing by the exposure.  
The difference between the residual
fluxes for the on- and off-pulse maps shows the expected distribution
for the EGRET PSF with a source coincident with PSR~\psra.

\subsection{PSR~\psrb} \label{sec:1105}

\subsubsection{Timing Analysis} \label{1105time}

\begin{figure}
\plotone{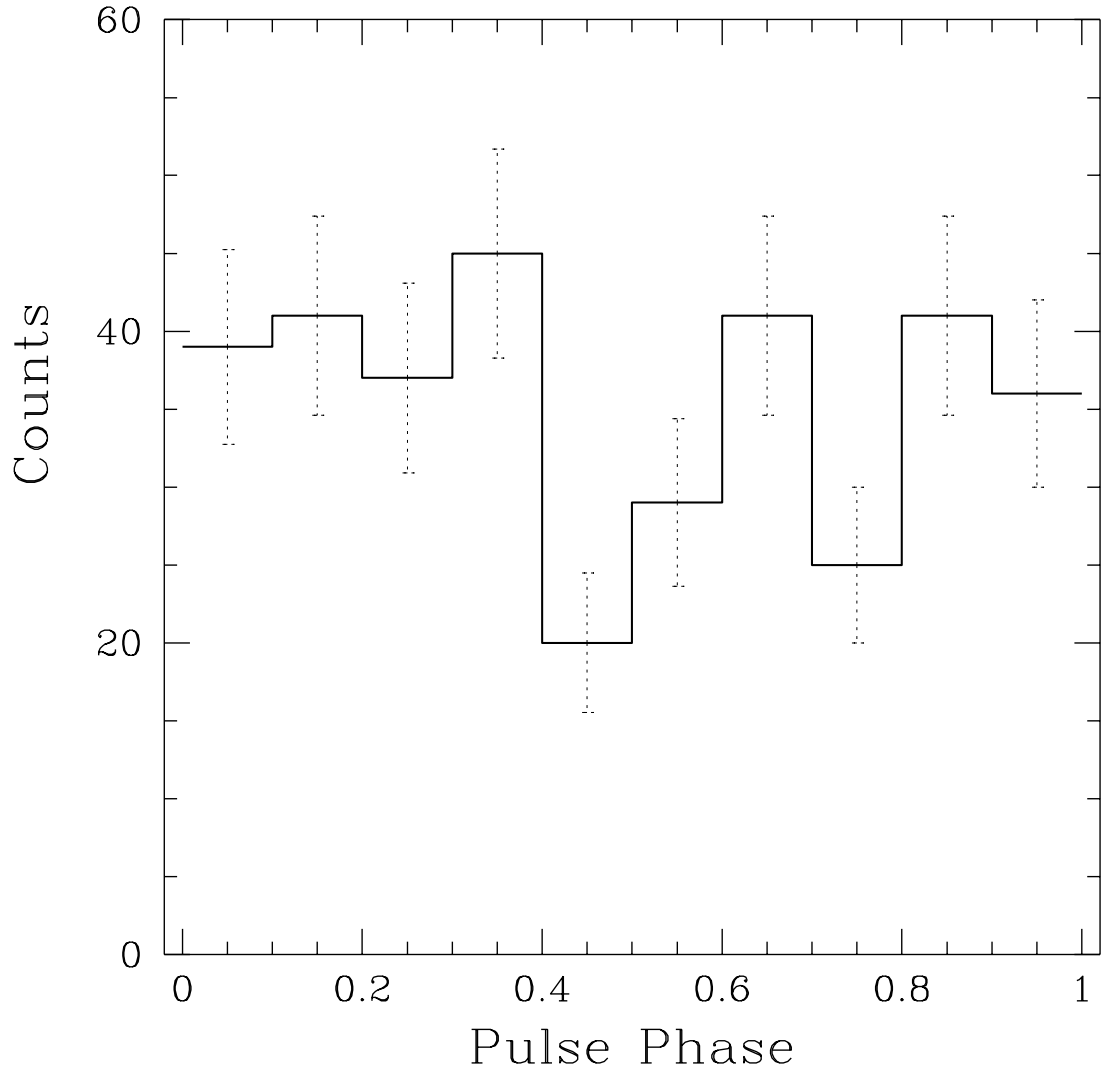}
\caption{Folded EGRET light curve for PSR~J1105$-$6107 for $E>400$~MeV.}
\label{fig:1105prof}
\end{figure}

In Figure~\ref{fig:1105prof}, we show the folded light curve for
PSR~\psrb\ for energies above 400~MeV.  The H
statistic is 3.6, which has a probability of 0.24 of having occurred by
chance.  We thus find no evidence for \gray\ pulsations from
PSR~\psrb.  We also tried folding photons from within a cone
encircling the source direction that contains 99\% of the
events; again no evidence for pulsations was found.  We also searched
for pulsations in different energy ranges to take advantage of the
expected increased pulsed fraction with increasing low energy cutoff,
but found no significant signals.  
Finally, we considered data weighted according to the
algorithm described in \S\ref{sec:1046}, but again found no
significant pulsations.  The 99\% confidence level upper limit on the
pulsed fraction of \grob\ at the PSR~\psrb\ spin period for $E> 400$~MeV,
using the H test and assuming a duty cycle of 0.5 is 0.33.  For a duty
cycle of 0.1, it is 0.13.  For $E>100$~MeV photons, the analogous
limits are 0.09 and 0.04 for duty cycles 0.5 and 0.1, respectively.

\section{Discussion} \label{sec:discuss}

\subsection{PSR~\psra}

The folded \gray\ profile for PSR~\psra\ and the spatial analysis of
the \gray\ data provide evidence for the association between
PSR~\psra\ and \groa.  However additional support for this conclusion
is desirable, because of the difficulty in interpreting the pulse
significance as a function of energy and the drop in significance when
photon weighting is used.  We consider here evidence that provides
additional support for the association between PSR~\psra\ and \groa.

\subsubsection{ASCA X-Ray Observations of the \groa\ Field} \label{sec:asca}

Additional evidence in favor of associating
PSR~\psra\ with \groa\ are the results of ASCA X-ray observations of the
\groa\ field, which are described in a companion paper (\cite{pkg99}).
There, archival ASCA observations are presented that cover the
full 95\% confidence error circle of the counterpart to \groa\ as
determined by \nocite{hbb+99} Hartman et al. (1999) in the recently
published 3EG catalog (where the source name is 3EG J1048$-$5840).  As
Pivovaroff et al. show, the only X-ray source within the 95\%
confidence contour of the \gray\ source is coincident with PSR~\psra\
within the $<1'$ X-ray positional uncertainty.  
The source is unpulsed and is likely to be a synchrotron nebula.
However, there is an additional nearby
hard-spectrum X-ray source outside the 95\% confidence contour (but
within the 99\% contour) that could also be the counterpart, a
possibility we cannot rule out without an improved \gray\ source
position.  Nevertheless, the most likely counterpart from the X-ray
observations is PSR~\psra.

\subsubsection{Spectrum and Energetics}
\label{sec:spectrum}

Merck et al. (1996) \nocite{mbd+96} found that \groa\ had a spectral
index of 2.0$\pm$0.1, and noted that the \gray\ source had properties
consistent with the known \gray\ pulsar population, as well as the
positional coincidence with PSR~\psra.  Bertsch et al. (in preparation)
find a spectral index of $1.97 \pm 0.09$ for the 3EG counterpart of
\groa, 3EG J1048$-$5840, and an integrated flux for $E>100$
MeV of $(62 \pm 7) \times 10^{-8}$~photon cm$^{-2}$~s$^{-1}$ (R.
Hartman, personal communication).  We have done an independent
pulse-phase-resolved spectral analysis by considering the fluxes and
spectra for peaks 1 and 2 separately, using \grays\ arriving during the
off-peak 2 interval as a background estimate.  We find spectra for both
peaks that are consistent with that reported above as well as with each
other.

Using the results of the on-pulse spatial analysis
(\S\ref{sec:1046spatialres}) together with the reported 3EG point
source flux (\cite{hbb+99}) and spectrum (Bertsch et al. 1999), we
derive a pulsed flux for $E>400$~MeV of $(2.5 \pm 0.6) \times
10^{-10}$~erg~cm$^{-2}$~s$^{-1}$, which corresponds to an efficiency
for conversion of spin-down luminosity into pulsed $E>400$~MeV
$\gamma$-rays of 0.011$\pm$0.03 for 1~sr beaming, assuming $d=3$~kpc.
Using the derived total pulsed fraction
(\S\ref{sec:1046time}), the total number of counts and the exposure
yields consistent results.  The observed flux is consistent with
previously published upper limits (Thompson et al. 1994).

\subsubsection{Absence of Variability}

McLaughlin et al. (1996) \nocite{mmct96} quantified the variability of
EGRET sources and showed that known \gray\ pulsars are not variable,
in contrast to all known \gray-loud active galaxies for which accurate
\gray\ flux measurements have been made.  They further analyzed the
variability of the unidentified EGRET sources and concluded that
\groa\ is non-variable, consistent with its identification as an
energetic rotation-powered pulsar.  An analysis of the larger data set
used to produce the 3EG catalog reveals no evidence for variability
from the source, consistent with its identification as a
rotation-powered pulsar (Hartman et al. 1999).

\subsubsection{Pulse Profile Morphology and Relative Radio Phase}

The morphology of the \gray\ profile shown in
Figure~\ref{fig:1046prof} lends some support to its interpretation as
pulsations from PSR~\psra.  Note that in the profile, the difference
in height between the two peaks is not statistically significant, and
the bin resolution is optimal.  The two peaks, having separation
0.36$\pm$0.13 in pulse phase, is reminiscent of the \gray\ profile of the
Vela pulsar (e.g. \cite{kab+94}), which has comparable spin-down
parameters.  Similar \gray\ pulse profiles are also observed for the
Crab pulsar, Geminga, and PSR B1951+32 (Thompson 1996 and references
therein).  Hard X-ray observations of the Vela pulsar reveal a pulse
profile that is similar to that at \gray\ energies (\cite{sh98}); this
suggests that hard X-ray observations of PSR~\psra\ could confirm its
identification as a source of pulsed \grays.  It must be kept in mind,
however, that the profile of other well-established high-energy
$\gamma$-ray pulsars is not as clearly double peaked.  
For example, PSR~B1706$-$44, whose spin-down parameters most closely resemble
those of PSR~\psra, has a $\gamma$-ray profile whose morphology could be
interpreted, within the limited statistics, as either
single or double peaked with small peak separation.
Therefore the double peaked profile in Figure~1 does not
lend unambiguous support to the proposed association.

The \gray\ pulse's
offset from the single radio pulse (Fig. \ref{fig:1046prof})
is similar to those seen in the known high-energy \gray\ pulsars.  The
offset in time between the peak of the radio pulse and the approximate
centroid of the first of the two \gray\ peaks shown in
Figure~\ref{fig:1046prof} is 0.19$\pm$0.12 in pulse phase, with the
uncertainty dominated by the poor profile statistics.
The phase offset to the
approximate midpoint between the two apparent pulses is 0.37$\pm$0.13.
The uncertainty in these numbers due to the uncertainty of 
0.01~pc~cm$^{-3}$ in the
single-epoch DM measurement (\S\ref{sec:intro}) is negligible.
This offset is similar to those of the other \gray\ pulsars having
similar \gray\ pulse morphologies, namely the Crab (considering the
radio precursor as the analog of the PSR~\psra\ radio pulse), Vela,
and PSR B1951+32 (\cite{tho96b} and references therein).
These show an approximate empirical trend of radio--\gray\ phase
offset that increases
with characteristic age (e.g. \cite{tho96b}).  The observed
radio/\gray\ offset for PSR~\psra\ is in agreement with this
trend, which predicts an offset in the approximate range 0.32--0.38.

\subsubsection{Previous Discussion of the PSR~\psra\ / \groa\ Association}

In addition to Thompson et al. (1994) noting a hint of a \gray\
pulsation from PSR~\psra, Fierro (1995) \nocite{fie95} discussed a possible
association between the pulsar and \groa\ in detail.  He noted the
positional coincidence as well as the plausibility of the implied
spectral and energy properties of the putative \gray\ pulsar.  He also
noted the absence of variability in the source.  
Yadigaroglu \& Romani (1997) \nocite{yr97} used a
statistical test that quantifies the probability for positional
coincidence of unidentified EGRET sources with nearby young objects.
They identified the PSR~\psra\ / \groa\ association using this
analysis.
Zhang \& Cheng (1998) \nocite{zc98} argued that the PSR~\psra\ /
\groa\ association was plausible on the basis of predictions from a
version of the outer gap model of \gray\ production in pulsar
magnetospheres.

\subsubsection{Implications}

If the association between PSR~\psra\ and \groa\ is correct, then this
is only the eighth pulsar to show evidence for high-energy \gray\
pulsations.  The poor statistics in the folded profile and the
presence of significant off-pulse emission preclude detailed
analysis of the \gray\ emission properties.  Nevertheless, the
pulsar's apparent \gray\ properties resemble those of other known
\gray\ pulsars, particularly the Crab and Vela pulsars and 
PSR B1951+32 (radio and \gray\ pulse
morphologies and phase offset).
PSR~\psra\ further shares a
property common to all other known ``older'' \gray\ pulsars (that is,
excluding the Crab and PSR B1509$-$58), namely that its maximum
luminosity appears to be in the high-energy \gray\ range.  By
contrast, Pivovaroff et al. (1999) show that PSR~\psra's efficiency for
conversion of $\dot{E}$ into X-rays in the ASCA range is only $\sim 1
\times 10^{-4}$.

We now consider what the implications of the possible \gray\
pulsations from PSR~\psra\ are for models of \gray\ emission from
rotation-powered pulsars.  If the pulsations are real, the observed
phase offset of the radio pulse is difficult to understand in the
context of models in which the \grays\ are produced by particles which
are accelerated along open field lines near the neutron star by
electric fields (e.g. \cite{dh82,ds94}).  In these models, the radio
emission is thought to arise in the same polar cap region, so the
radio pulse should be observed between the \gray\ peaks, which result
from emission from a hollow cone.  The conal geometry results from the
smaller radius of curvature of magnetic field lines near the polar cap
rim (see Harding 1996 for a review). \nocite{har96b} This radio/\gray\
phase problem has been noted for other \gray\ pulsars as well,
e.g. Vela (\cite{dh96}); the results reported here argue that the
problem is general.  Also, emission beams in the polar cap model tend
to be narrow and aligned with the magnetic and spin axes
(e.g. \cite{sd94,hm98}); this can pose a problem for the pulsar birth
rate.  However, the radio phase offset problem, as argued by Harding
\& Muslimov (1998), \nocite{hm98} seems general to both the polar cap
\gray\ emission model as well as to geometric models of thermal X-ray
pulsations, while the relative phases of the latter two are
consistent.  Polar cap models therefore make a testable prediction
regarding the phase of the peak of any thermal X-ray emission that may
one day be detected from PSR~\psra.

The observed \gray\ pulse morphology and radio phase offset for
PSR~\psra\ are, by contrast, well explained in the outer gap model.
In this model, the \gray\ emission is produced in the charge-depleted
region between the closed magnetic field lines and the velocity of
light cylinder (e.g. \cite{chr86a,cr94,ry95,zc98}).  The double peaked
\gray\ pulse morphology in this model arises from a single pole's
outer gap that has uniform emissivity along its field lines but in
which the radiation is subject to significant relativistic aberration
and time-of-flight delays through the magnetosphere; the expected
pulse morphology for most viewing geometries is double peaked, as long
as the magnetic inclination is not small.  The morphology and phase
are also consistent with the geometry proposed by Manchester (1996)
\nocite{man96} in which the radio emission is produced in the outer
gap as well.  The outer gap model predicts bridge emission
between the two \gray\ peaks, whose outer edges should be sharp; this
could be tested with an improved \gray\ pulse profile.  This model
results in large beaming fractions that are consistent with the pulsar
birth rate; this implies that most, if not all, of the unidentified
EGRET low-latitude sources that have constant fluxes are young
rotation-powered pulsars (\cite{yr97}), although most of them will not
be detectable in radio waves from Earth because of smaller radio
beaming fractions.  We note, however, that recently reported very
tentative evidence for \gray\ emission from the millisecond pulsar
PSR~J0218+4232 (\cite{khv+98}) could pose a problem for the outer gap
model, which predicts that millisecond pulsars should not be
detectable in high-energy \grays\ (\cite{ry95}).

\subsubsection{Unpulsed \gray\ Emission from PSR~B1046$-$58?}

The spatial analysis of the EGRET data presented in
\S\ref{sec:1046spatialres} showed evidence for off-pulse
emission consistent with coming from a point source
coincident with the pulsar.  For $E>400$~MeV,
this emission has flux 
$(1.9 \pm 0.6) \times 10^{-10}$~erg~cm$^{-2}$~s$^{-1}$,
which corresponds to $\sim 10$\% of $\dot{E}$, for
isotropic emission.  The poor spatial resolution
of EGRET obviously precludes a firm association between the pulsar and this
unpulsed emission, although we note that off-pulse \gray\ emission
has been seen for the Crab, Vela and Geminga pulsars, with that
of the Crab possibly originating in the surrounding nebula (\cite{fmnt98}).

\subsection{PSR~\psrb}

Since we find no evidence for \gray\ pulsations for PSR~\psrb, we set
an upper limit on the fraction of $\dot{E}$ that is converted into
high-energy \grays\ beaming in our direction.  The counterpart to
\grob\ in the 3EG catalog, 3EG J1102$-$6103, has flux $(32.5 \pm 6.2)
\times 10^{-8}$~photons~cm$^{-2}$~s$^{-1}$ for $E>100$~MeV, and photon
index $2.47 \pm 0.21$ (\cite{hbb+99}).  Using these numbers, we infer
with 99\% confidence that PSR~\psrb\ converts less than $8 \times
10^{-11}$~erg~cm$^{-2}$~s$^{-1}$ and $3 \times
10^{-11}$~erg~cm$^{-2}$~s$^{-1}$ for duty cycles of 0.5 and 0.1,
respectively.  These limits correspond to efficiencies for conversion
of spin-down luminosity of less than 0.014 and 0.006 for 1~sr beaming
and assuming a distance of 7~kpc, for duty cycles 0.5 and 0.1,
respectively.  These limits are not constraining on models of \gray\
emission.

\section{Conclusions} \label{sec:concl}

We have found evidence for \gray\ emission from the young, energetic
radio pulsar PSR~\psra, which, if correct, establishes a counterpart
for the previously unidentified high-energy \gray\ source \groa\ (GEV
J1046$-$5840, 3EG J1048$-$5840).  The evidence can be summarized as
follows.  (i) PSR~\psra\ has the ninth highest value of $\dot{E}/d^2$
of the known radio pulsars, with six of the eight higher slots
occupied by known \gray\ pulsars.  This suggests that PSR~\psra\ is
likely to be detectable in \grays.  (ii) The radio pulsar is spatially
coincident with the unidentified EGRET source \groa.  (iii) The folded
light curve for $E> 400$~MeV, obtained using contemporaneous radio
ephemerides, is characterized by an H-test statistic 22.5; the chance
probability of this value or higher having occurred is $1.2 \times
10^{-4}$.  This number should be multiplied by a factor of two for the
two pulsars we searched.  (iv) A spatial likelihood analysis of on-
and off-pulse data reveals a point source coincident with the radio
pulsar position whose significance is much greater on-pulse than off
(10.2$\sigma$ versus 5.8$\sigma$).  (v) The only hard X-ray point
source detected in the 95\% EGRET error box is spatially coincident
with the pulsar (Pivovaroff et al. 1999).  (vi) The \gray\ light curve
morphology resembles those of other the known \gray\ pulsars (the Crab
and Vela pulsars, Geminga, and PSR~B1951+32), namely, two narrow peaks
separated by $\sim$0.4 in pulse phase.  (vii) The inferred
radio/\gray\ phase offset is consistent with trends seen for other
pulsars.  The implied efficiency for conversion of spin-down energy into
$E>400$~MeV \grays, 0.011$\pm$0.003 for $d=3$~kpc
and 1~sr beaming, is consistent with those seen for
other pulsars.  We also found evidence for spatially unresolved
unpulsed \gray\ emission from the pulsar position, as has been
reported for the Crab, Vela and Geminga pulsars.

If the association between PSR~\psra\ and \groa\ emission exists, it should be
confirmed by GLAST, the planned Gamma-ray Large Area Space Telescope,
which will provide significantly higher angular resolution and more
detector area than EGRET.  It may be possible to confirm the \gray\
pulsations by detecting magnetospheric X-ray pulsations from
PSR~\psra\ at hard X-ray energies.  The similarity of this pulsar's
rotational parameters and high-energy \gray\ efficiency to those of
PSR~B1706$-$44 suggest TeV emission might be detectable.  Thermal
emission from the initial cooling of this young neutron star may also
be detectable.

If the association between PSR~\psra\ and \groa\ holds, that seven of
the top nine pulsars ranked by $\dot{E}/d^2$ are \gray\ pulsars (with
one of the two remainders not having been studied at \gray\ energies
-- see Table~1) implies at the very least that the radio and \gray\
beams are aligned closely, or that \gray\ beams are wide.  In fact,
the morphology of the observed \gray\ pulse and the phase offset of
the radio pulse provide support for the outer gap model of \gray\
emission (\cite{ry95} and references therein).  This model predicts that an
improved \gray\ profile should find bridge emission between the pulses
whose outer edges should be sharp.  In this model, most or all of the
unidentified EGRET sources that have constant fluxes are young
rotation-powered pulsars like PSR~\psra, though most radio beams will
not be detectable from Earth due to small radio beaming fractions.

Our analysis of EGRET data from the direction of PSR~\psrb\
finds no evidence for \gray\ pulsations.  This pulsar converts
at most 0.014 (99\% confidence) of its spin-down power to $E>100$~MeV
\grays\ beamed in our direction, assuming 1~sr beaming, a duty cycle
of 0.5, and a distance to the source of 7~kpc.  This limit is
not constraining on models.

\section*{Acknowledgements}

We are grateful to Dave Thompson for many helpful conversations, and
Robert Hartman for communication of Third EGRET Catalog numbers prior
to publication.  We also thank 
Russell Pace for help with the Parkes observations.  This research has
made use of data obtained through the High Energy Astrophysics Science
Archive Research Center Online Service, provided by the NASA/Goddard
Space Flight Center.  This work was supported by NASA grants NAG
5-3178 and NAG 5-3683 to VMK, and NASA LTSA grant NAG 5-3384 to JRM.
We thank Joe Fierro for inspiration and assistance in the initial
undertaking of this project.


\clearpage

\clearpage

\clearpage
 
\clearpage

\begin{deluxetable}{lccccccc}
\tablecaption{Summary of Known High-Energy \gray\ Pulsars and
PSRs~\psra\ and \psrb, ordered by $\tau$.}
\label{ta:psrs}
\tablehead{
\colhead{Pulsar\tablenotemark{a}} & \colhead{$P$} &  \colhead{$\log \tau$\tablenotemark{b}} & \colhead{$\log B$\tablenotemark{c}} & \colhead{$\log \dot{E}$\tablenotemark{d}} & 
 \colhead{$d$\tablenotemark{e}} &  \colhead{$\dot{E}/d^2$}\tablenotemark{f} &
\colhead{$\eta_{\gamma}$\tablenotemark{g}}  \\ 
\colhead{} & \colhead{ms} & \colhead{yr} & \colhead{G} &
 \colhead{erg~s$^{-1}$~cm$^{-2}$} & \colhead{kpc} & \colhead{rank} & \colhead{\%}  }

\startdata
Crab & 33 & 3.1 & 12.6 & 38.7 & 2.0 & 1 &  0.013$\pm$0.006               \nl
Vela & 89 & 4.1 & 12.5 & 36.8 & 0.5 & 2 & 0.18$\pm$0.07
\nl
B1706$-$44 & 102 & 4.2 & 12.5 & 36.5 & 1.8 & 4 & 0.72$\pm$0.47 \nl
{\bf B1046$-$58} & 124 & 4.3 & 12.5 & 36.3 & 3.0 & 9 & 1.1$\pm$0.3\nl
{\bf J1105$-$6107} & 63 & 4.8 & 12.0 & 36.4 & 7.1 & 23 & $<1.4$ \nl
B1951+32 & 39 & 5.0 & 11.7 & 36.6 & 2.4 & 6 & 0.26$\pm$0.17    \nl
B0656+14 & 385 & 5.0 & 12.7 & 34.6 & 0.8 & 20 & $\sim$0.1     \nl
Geminga & 237 & 5.5 & 12.2 & 34.5 & 0.157$^{+0.059}_{-0.034}$ & 3 & 1.61$^{+0.22}_{-0.08}$   \nl
B1055$-$52 & 197 & 5.7 & 12.0 & 34.5 & 1.5 & 39 & 6--13    \nl

\enddata
\tablenotetext{a}{Pulsars in bold are the subject of this paper.}
\tablenotetext{b}{$\tau \equiv P/2\dot{P}$}
\tablenotetext{c}{$B \equiv 3.2\times10^{19}(P\dot{P})^{1/2}$ G}
\tablenotetext{d}{$\dot{E} \equiv 4\pi^2 I \dot{P}/P^3$}
\tablenotetext{e}{Distances from Taylor \& Cordes (1993)
\nocite{tc93} except for Geminga which is from Caraveo et
al. (1996).\nocite{cbmt96} }
\tablenotetext{f}{Ranked 5th is PSR B1509$-$58, detected in low-energy
\grays\ (\cite{umw+93}); 7th is the millisecond pulsar PSR
J0437$-$4715; 8th is PSR J1617$-$5055, which was only
recently discovered (\cite{tkt+98,kcm+98}).}
\tablenotetext{g}{For $E>100$~MeV and 1~sr beaming, except for
PSR~B1046$-$58
which is for $E>400$~MeV.  Uncertainties
include nominal distance uncertainties.  Numbers are from Fierro
(1995) except for Geminga, for which we have used the updated
distance, PSR B0656+14, which is from Ramanamurthy et al. (1996),
\nocite{rfk+96} and PSR B1055$-$52, which is from Thompson et al. (1999).
\nocite{tbb+99}  For PSR J1105$-$6107, the
3$\sigma$ upper limit for duty cycle 0.5 is quoted.}

\label{ta:psrs}
\end{deluxetable}

\clearpage
\begin{deluxetable}{ccccc}
\tablecaption{Summary of EGRET Observations for PSR~B1046$-$58.} 
\tablehead{\colhead{VP\tablenotemark{a}} & \colhead{MJD Range} &
\colhead{$\Delta\theta$\tablenotemark{b}} & \colhead{$N(\theta_{67})$} &
\colhead{Exposure} \\ & & \colhead{$^{\circ}$} & & }
\startdata
0007 &	48386--48392 &	21.1 &	614 &	0.52 \\
0060 &	48463--48476 &	31.2 &	279 &	0.54 \\
0080 &	48490--48504 &	25.2 &	1032 &	1.01 \\
0120 &	48546--48560 &	31.3 &	235 &	0.49 \\
0140 &	48574--48588 &	2.7 &	3252 &	3.08 \\
0170 &	48617--48631 &	32.4 &	140 &	0.45 \\
0230 &	48700--48714 &	34.8 &	30 &	0.14 \\
0320 &	48798--48805 &	22.5 &	289 &	0.40 \\
2080 &	49020--49027 &	28.1 &	163 &	0.18 \\
2150 &	49078--49083 &	32.4 &	52 &	0.08 \\
2170 &	49089--49097 &	32.4 &	34 &	0.12 \\
2300 &	49195--49198 &	11.1 &	335 &	0.45 \\
2305 &	49198--49202 &	8.7 &	416 &	0.60 \\
3010 &	49216--49223 &	24.1 &	361 &	0.40 \\
3140 &	49355--49368 &	16.8 &	775 &	1.09 \\
3150 &	49368--49375 &	16.8 &	381 &	0.60 \\
3160 &	49375--49384 &	28.7 &	157 &	0.24 \\
3385 &	49595--49615 &	24.1 &	649 &	0.87 \\
4020 &	49643--49650 &	23.5 &	156 &	0.24 \\
4025 &	49650--49657 &	19.8 &	195 &	0.40 \\
4150 &	49818--49832 &	27.1 &	261 &	0.36 \\
4240 &	49908--49923 &	30.9 &	65 &	0.24 \\
5220 &	50245--50248 &	2.7 &	136 &	0.33 \\
5310 &	50359--50371 &	3.7 &	366 &	1.16 \\
6270 &	50693--50700 &	1.6 &	234 &	1.01 \\
6300 &	50714--50728 &	1.6 &	437 &	1.89 \\
\enddata
\tablenotetext{a}{EGRET Viewing Period.}
\tablenotetext{b}{Angular offset of pulsar from center of EGRET field
of view.}
\label{ta:1046obs}
\end{deluxetable}

\clearpage
\begin{deluxetable}{ccccc}
\tablecaption{Summary of EGRET Observations of PSR~J1105$-$6107.} 
\tablehead{ \colhead{VP\tablenotemark{a}} & \colhead{MJD Range} &
\colhead{$\Delta\theta$\tablenotemark{b}} & \colhead{$N(\theta_{67})$} &
\colhead{Exposure} \\ & & \colhead{$^{\circ}$} & & }
\startdata
2300 &	49195--49198 &	14.0 &	293 &	0.38 \\
2305 &	49198--49202 &	11.8 &	412 &	0.52 \\
3010 &	49216--49223 &	27.0 &	245 &	0.30 \\
3140 &	49355--49368 &	13.7 &	826 &	1.32 \\
3150 &	49368--49375 &	13.7 &	420 &	0.71 \\
3160 &	49375--49384 &	26.9 &	213 &	0.27 \\
3385 &	49595--49615 &	27.0 &	430 &	0.67 \\
4020 &	49643--49650 &	20.4 &	201 &	0.32 \\
4025 &	49650--49657 &	16.7 &	226 &	0.31 \\
4150 &	49818--49832 &	28.0 &	213 &	0.35 \\
4240 &	49908--49923 &	28.9 &	114 &	0.28 \\
5220 &	50245--50248 &	5.0 &	128 &	0.32 \\
5310 &	50359--50371 &	6.7 &	304 &	0.99 \\
6270 &	50693--50700 &	3.6 &	205 &	0.96 \\
6300 &	50714--50728 &	3.6 &	419 &	1.81 \\
\enddata 
\tablenotetext{a}{EGRET Viewing Period.}
\tablenotetext{b}{Angular offset of pulsar from center of EGRET field
of view.}
\label{ta:1105obs} 
\end{deluxetable}

\end{document}